\documentclass{emulateapj}
\usepackage{apjfonts}
\usepackage{epsfig}
\usepackage{natbib}

\newlength{\colwidth}

\newcommand{\be}{\begin{equation}}
\newcommand{\ee}{\end{equation}}
\newcommand{\bee}{\begin{eqnarray}}
\newcommand{\eee}{\end{eqnarray}}

\newcommand{\kms}{km s$^{-1}$}

\newcommand{\msol}{\hbox{${\rm M}_\odot$}}

\newcommand{\mvir}{\hbox{$M_{\rm vir}$}}

\newcommand{\vmax}{v_{\rm max}}
\newcommand{\vacc}{v_{\rm acc}}

\newcommand{\LCDM}{$\Lambda$CDM }

\newcommand{\hinv}{h^{-1}}
\newcommand{\mpc}{\rm{Mpc}}
\newcommand{\hmpc}{h^{-1}\rm{Mpc}}
\newcommand{\hMpc}{h^{-1}\rm{Mpc}}
\newcommand{\hkpc}{h^{-1}\rm{kpc}}

\newcommand{\N}{$N_{subs}$}
\newcommand{\PN}{$P(N_{subs})$}
\newcommand{\NS}{$N_{sats}$}
\newcommand{\PNS}{$P(N_{sats})$}

\newcommand{\lsim}{\lower.5ex\hbox{\ltsima}}
\newcommand{\gsim}{\lower.5ex\hbox{\gtsima}}
\newcommand{\ltsima}{$\; \buildrel < \over \sim \;$}
\newcommand{\gtsima}{$\; \buildrel > \over \sim \;$}

\usepackage{color}
\newcommand{\mike}[1]{#1}

\definecolor{purple}{RGB}{160,32,240}

\shortauthors{BUSHA ET AL.}
\shorttitle{Statistics of Satellite Galaxies of the Milky Way}
\begin{document}

\title{Statistics of Satellite Galaxies Around Milky Way-Like Hosts}

\author{Michael T. Busha\altaffilmark{1,2}, Risa
  H. Wechsler\altaffilmark{1,3}, Peter S. Behroozi\altaffilmark{1,3},
  Brian F. Gerke\altaffilmark{1,3}, \\Anatoly
  A. Klypin\altaffilmark{4}, and Joel R. Primack\altaffilmark{5}}

\altaffiltext{1}{Kavli Institute for Particle Astrophysics and Cosmology, \\
Department of Physics, Stanford University, Stanford, CA, 94305, USA\\
{\tt mbusha@slac.stanford.edu, rwechsler@stanford.edu}}
\altaffiltext{2}{Institute for Theoretical Physics, University of
  Zurich, 8057 Zurich, Switzerland}  
\altaffiltext{3}{SLAC National Accelerator Laboratory, Menlo Park, CA, 94025, USA}
\altaffiltext{4}{Astronomy Department, New Mexico State University,
  Las Cruces, NM, 88003, USA}
\altaffiltext{5}{Department of Physics, University of California, Santa Cruz, CA 95064, USA}

\begin{abstract}

 We calculate the probability that a Milky-Way-like halo in the
 standard cosmological model has the observed number of  Magellanic
 Clouds (MCs). The statistics of the number of MCs in the \LCDM model
 are in good agreement with observations of a large  sample of SDSS
 galaxies.  Under the sub-halo 
  abundance matching assumption of a relationship with small scatter
  between galaxy $r$-band luminosities and halo internal velocities
  $\vmax$, we make detailed comparisons to similar measurements using
  SDSS DR7 data by \cite{Liu10}.  Models and observational data give very similar
  probabilities for having zero, one, and two MC-like satellites.  In both cases,
  Milky Way-luminosity hosts have just a $\sim 10$\% chance of hosting
  two satellites similar to the Magellanic Clouds.  In
  addition, we present a prediction for the probability for a host
  galaxy to have \NS\ satellite galaxies as a function of the
  magnitudes of both the host and satellite. This probability and its
  scaling with host properties is significantly different from that of
  mass-selected objects because of scatter in the mass-luminosity
  relation and because of variations in the star formation efficiency
  with halo mass.

\end{abstract} 


\section{Introduction}

Understanding the satellite population of our own Milky Way galaxy is
one of the outstanding problems in galaxy evolution studies today.
Because these objects have weak gravitational potential wells, it is
quite likely that different physical processes played a driving role
in determining how such galaxies formed and evolved than for brighter,
more massive objects.  One manifestation of this is the well-known
``missing satellites problem,'' which refers to the fact that the
number of satellite galaxies predicted from simulations appears to be
much higher than the number actually observed around the Milky Way
\citep{Klypin99, Moore99b}.  One common way to address this problem
has been to run simulations of individual objects using extremely high
mass resolution.  To date, roughly 10 of these ultra-high resolution
N-body simulations have been run, which typically resolve individual
Milky Way mass halos with upwards of $10^9$ particles \citep{Madau08a,
  Springel08, Stadel08}.

The Magellanic Clouds (MCs) are a pair of well studied satellites of
the Milky Way (MW) in the Southern Hemisphere with magnitudes $M_V =
-18.5$ and -17.1 for the Large and Small Magellanic Clouds (LMC and SMC).  While their large angular size makes
detailed mass measurements somewhat difficult, observations constrain
their maximum circular velocities to be around $\vmax \sim 60$ \kms
\citep{vandenBergh00, VanDerMarel02, Stanimirovic04}.  Unlike the
missing satellite problem, one common feature of these ultra-high
resolution simulations is that they typically under-predict the number
of massive satellites compared to the MCs in the MW \citep[see, e.g.,
Figure 5 in][]{Madau08a}.  Indeed, most of these high resolution halos
contain no objects of similar mass to either of the two MCs.  While
this should hardly be taken as a serious problem since the handful of
high resolution simulated halos does not constitute a statistical
sample, it is important to understand how typical or atypical the
Milky Way is, and whether or not not there is an ``extra satellites''
problem at high satellite masses within the modern Lambda Cold Dark
Matter ($\Lambda$CDM) paradigm.

Because the MCs are relatively nearby and have been studied with
high-precision instruments such as HST for some time, detailed
3-dimensional measurements of their velocities have been made
\citep{Kallivayalil06a, Kallivayalil06b}.  Some of these studies have
indicated that the Magellanic Clouds may be on their first orbit
around the Milky Way \citep{Besla07, Busha10c}; this may make such an
extra satellite problem less worrisome if the presence of the
Magellanic Clouds is a transient event.  Regardless of the dynamical
state of the Clouds, it is an important test of galaxy formation to
understand the likelihood for MW-like systems to host massive
satellite galaxies.

Recent developments have afforded us the possibility to address this
problem in more detail with both observations and theoretical models.
From the observational side, wide area surveys such as the Sloan
Digital Sky Survey \citep[SDSS][]{York00,Abazajian09} have given us
the ability to probe galaxy content for a large number of Milky
Way-magnitude objects \citep{Chen06, James10, Liu10}. 

The main theoretical effort has been in populating dark matter halos
with galaxies using semi-analytic methods and hydrodynamic
simulations.  \citet{Koposov09} used a number of toy models to add
galaxy properties to dark matter halos generated using a combination
of Press-Schechter theory with semi-analytic models for tracking
subhalo orbits \citep{Zentner05}.  They found that it was very
difficult to model objects as bright as Magellanic clouds without
allowing for an extremely high star formation efficiency.  This
finding extended the underprediction of high-mass subhalos to an
underprediction of luminous satellites, implying that the MW-MCs
system is unusual.  Similarly, \citet{Okamoto10} explored a range of
feedback models to add galaxies to some of the high resolution
Aquarius halos \citep{Springel08}.  It was again difficult to readily
reproduce halos with luminosities as bright as the MCs.  Having looked
in detail at a handful of simulated objects, however, this work
indicates that there may be significant halo-to-halo scatter in the
number of such massive objects.  The idea of intrinsic scatter in the
subhalo population was expanded on in \cite{Ishiyama09}.  While this
work did not concentrate specifically on MC-like subhalos, they did
consider the range in number of subhalos with
$v_{max,sub}/v_{max,host} > 0.1$.  This work showed an extremely large
variation (20-60) in the number of such massive subhalos a
galaxy-sized halos would host.

In contrast to the semi-analytic modeling just discussed,
\cite{Libeskind07} used hydrodynamic simulations to model the
luminosity functions for the satellites around MW-like host halos.
Their simulation identified 9 MW-like central galaxies and
found that they, on average, have 1.6 satellites brighter than $M_V =
-16$ and a third of them had satellites with luminosities comparable
to the LMC.

The recent Millennium-II and Bolshoi simulations
\citep{BoylanKolchin10,Klypin10} have, for the first time, allowed us
to probe cosmological volumes to understand the \LCDM predictions for
the satellite populations of MW-like halos.  The properties of
these simulations are summarized in Table 1.  \citet{BoylanKolchin10}
(hereafter BK10) quantified the likelihood for $10^{12} \msol$ halos
to host massive satellite galaxies in the Millennium-II simulation,
finding that subhalos similar to the MCs are quite rare.  In this
paper, we expand on this work by making similar measurements for the
Bolshoi simulation, which used WMAP7 cosmological parameters,
and using an abundance matching technique to make detailed comparisons
between the Bolshoi predictions and the measurements from
\cite{Liu10}.  Our goal is to understand just how well \LCDM
reproduces the statistical properties of bright satellites.

Note that this is the reverse question from the one that was asked in
a companion paper, \cite{Busha10c}.  That work assumed a satellite
population and asked what the implications were for the properties of
the host halo, including its mass.  Here, we assume a host halo mass
and ask about the implications for the subhalo population.  There is
no reason for both questions to give the same answer: while
\cite{Busha10c} showed that a halo which hosts two MC-like satellites
most likely has a mass near $1.2 \times 10^{12} \msol$, there is no
reason to assume that a typical $1.2 \times 10^{12}\msol$ halo will
have the MCs as satellites.

We begin by giving an overview of the Bolshoi simulation in \S
\ref{sec:sims}, and then investigate the properties of massive dark
matter satellites around dark matter host halos in \S \ref{sec:halos},
focusing on the mass ranges for hosts and satellites that are most
relevant to the MW system.  The analysis here is similar to that of
BK10.  In \S \ref{sec:luminosities}, we assign galaxy luminosities to
our suite of dark matter halos and extend the results for a sample
with similar selection cuts as for observations.  In this way, we are
able to make detailed comparisons to the observational work of
\cite[][hereafter L10]{Liu10} concerning the satellite population
around MW-magnitude galaxies.  Section \ref{sec:comparisons} gives the
results of this analysis --- see especially Figure
\ref{fig:nsats_obs_sim}.  Finally, in \S \ref{sec:generalproperties},
we expand this study to include the satellite population of a more
general distribution of hosts, and \S \ref{sec:conclusions} summarizes
our conclusions.  Throughout this paper, we adopt the convention $h =
0.7$ (the value that was used in the Bolshoi simulation) when
reporting values from either simulations or observations.

\section{Simulations} 
\label{sec:sims} 
We use the dark matter halos identified in the Bolshoi simulation
\citep{Klypin10, TrujilloGomez10}.  This simulation modeled a 250
$\hmpc$ comoving box using cosmological parameters similar to those
derived by WMAP7 \citep{Komatsu10}: $\Omega_m = 0.27$,
$\Omega_{\Lambda} = 0.73$, $\sigma_8 = 0.82$, $n=0.95$, and $h = 0.7$.
The simulation volume contains $2048^3$ particles, each with a mass of
$1.35 \times 10^8~\hinv\msol$ and was run using the ART code
\citep{Kravtsov97}.  180 snapshots from the simulation were saved and
analyzed.  One of the unique aspects of this simulation is the high
level of spatial resolution employed, allowing objects to be resolved
down to a physical scale of 1 $\hkpc$.  This gives us excellent
ability to track halos as they merge with and are disrupted by larger
objects, allowing us to track them even as they pass near the core of
the host halo.  A summary of these simulation parameters is presented
in Table \ref{table:simulations}.  Because we discuss our work in the
context of the BK10 results, we also present the same parameters for
the Millennium II simulation \citep{BoylanKolchin09}, on which the
BK10 results were based.

Halos and subhalos were identified using the BDM algorithm
\citep{Klypin97}.  The algorithm identifies maxima in the density field
and examines the neighboring region to identify bound particles.  In this
way, it treats both halos and subhalos identically. Subhalos are
just identified as objects living within the virial radius of a larger
objects.  Because of the high level of mass and spatial resolution,
BDM results in a halo catalog that is complete down to a maximum
circular velocity $\vmax = 50$ km s$^{-1}$, where  
$\vmax = {\rm max}\left(\sqrt{{GM(<r)}/{r}}\right). $ 
This corresponds to a virial mass of roughly $10^{10}\hinv\msol$.
When BDM halos are identified, the ids of their 50 most bound
particles are also stored to assist in producing merger trees.   

As discussed in Section \ref{sec:sham}, in order to assign galaxy
luminosities to dark matter halos, we need to track the histories of
dark matter substructures.  This is done using merger trees created
from the 180 snapshots of the Bolshoi simulation.  The detailed algorithm
for creating the merger trees is described in \cite{Behroozi10b}.
Briefly, the algorithm works by first linking halos across time steps
by tracking the 50 most bound particles of each halo.  Some
halos will not have any of their 50 most bound particles identified at
a later timestep (e.g., very massive objects in which the 50 most bound
particles change rapidly through stochastic processes), while
some will have their particles distributed to multiple halos.  The
algorithm corrects for this by running a simple N-body calculation on
the locations and masses of all halos in the simulation to
predict where each halo should wind up at the next time step.
Using this information, it is possible to link halos across multiple
time steps with very high accuracy. 
 
\begin{table} 
\caption{Comparison of the simulation parameters from Bolshoi (used for this work) with those of the Millennium II Simulation (used in BK10)}  
\begin{center} 
\begin{tabular}{ c c c } \hline
       & Bolshoi & MSII \\
      \hline \hline
      Box Size [Mpc] & 357 & 143 \\
      $N_P$ & $2048^3$ & $2160^3$ \\
      $M_P [\msol]$ & $1.9 \times 10^8$ & $9.8 \times 10^6$ \\
      $h$ & 0.7 & 0.73 \\
      $\Omega_M$ & 0.27 & 0.25 \\
      $\sigma_8$ & 0.82 & 0.9 \\
      Force Resolution [kpc] & 1.4 (proper) & 1.4 (Plummer-equivalent softening)\\
 
      \hline \end{tabular} 
\end{center}
\label{table:simulations}

\end{table}

\section{Satellite Statistics for Dark Matter Subhalos}
\label{sec:halos}

We begin by investigating the properties of dark matter satellites
around dark matter hosts in the Bolshoi simulation, focusing on the
mass range for hosts and satellites that is relevant to the MW system.
In \S \ref{sec:massdep}, we consider trends with host halo mass, and in \S
\ref{sec:env} we investigate trends with halo environment.  We then
consider the distribution of the satellite number (\S \ref{sec:pois}).
Finally, we extend this analysis to include a more observationally
relevant selection based on galaxy luminosities in \S
\ref{sec:luminosities}. 

Because the mass resolution of the Bolshoi simulation creates a halo
catalog complete down to 50 \kms, roughly equivalent to the lower
bound of the mass of the MCs \citep{VanDerMarel02, Stanimirovic04,
  Harris06}, we begin by measuring the probability distribution for
halos hosting \N\ subhalos with $\vmax$ larger than a given value.
This measurement is made by identifying the virial radius of all
distinct (non-subhalo) MW mass halos and counting the number of
objects internal to their virial radii.  We take the virial mass of
the Milky Way to be $\log(\mvir/\msol) = 12.08 \pm 0.12$ ($\mvir = 1.2
\times 10^{12} \msol$), the mass measured in \cite{Busha10c}, which is
consistent with most results in the literature \citep{Battaglia05,
  Dehnen06, Smith07, Xue08}.  This results in 36,000 MW analogs found
in the Bolshoi volume.  We present the resulting probability
distribution of satellite counts as the colored points in the left
panel of Figure \ref{fig:vmax_comp}.  Here, the different colors
represent different thresholds for satellite $\vmax$: red, green, and
blue represent all satellites with $\vmax > $ 50, 60, and 70 \kms.
Error bars were calculated using the bootstrap method and represent
95\% confidence intervals.

From Figure \ref{fig:vmax_comp}, we can immediately see that the
likelihood of hosting multiple satellites is somewhat low.  For satellites with $\vmax$ 
greater than 50 \kms, roughly 30\% of simulated hosts contain one or
more subhalos and just 8\% have two or more.  These numbers are in
excellent agreement with the work of BK10, who performed a similar
analysis on the Millennium II simulation, which was run using a
different N-body code in a WMAP1 cosmology and with a very different
subhalo identification algorithm.  The probability of hosting multiple
satellites drops precipitously with increasing satellite mass; e.g.,
the probability of a halo hosting two subhalos larger than $\vmax =
70$\kms{} is just 1\%.

For comparison, the black points in Figure \ref{fig:vmax_comp}
represent measurements from \cite{Liu10}, who measured the probability
distribution for finding MC-luminosity satellites within 250 kpc
around isolated MW-luminosity galaxies (the mean virial radius of our
sample).  We only plot these measurements out to \N\ = 3 because, for
a 250 kpc aperture, uncertainties from their background subtraction
become extremely large for higher (lower probability) values of \N.
Here, the agreement for satellites larger than 50 \kms is striking,
indicating that CDM can reproduce the statistics of MCs.  We must,
however, be careful about over-interpreting this result.  In
particular, their selection criteria are very different from our mass
selection.  We will return to this in \S 4.4 where we make a
comparison directly to similarly selected samples.

By comparison with the Bolshoi simulation, the existence of two
satellites with $\vmax$ larger than 50\kms\ makes the MW almost a
2$\sigma$ outlier.  Understanding whether anything other than random
chance is responsible for putting the MW in this slightly rare regime
motivates further investigation into which other properties may
correlate with satellite abundance.  In particular, we investigate the
correlation of the number of subhalos with host mass and environment.

\subsection{Mass dependence}
\label{sec:massdep}
\begin{figure*}[t]
  \plottwo{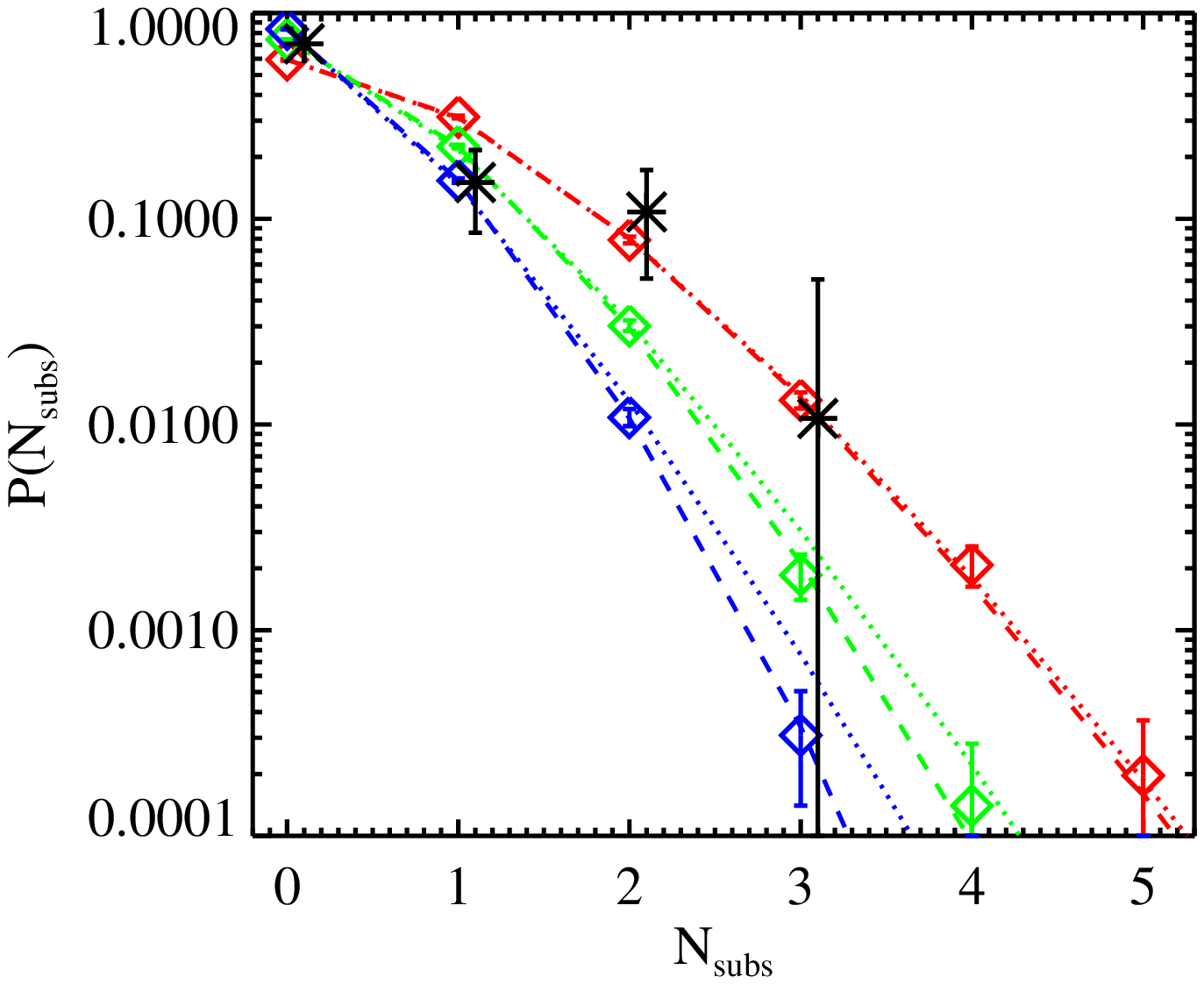}{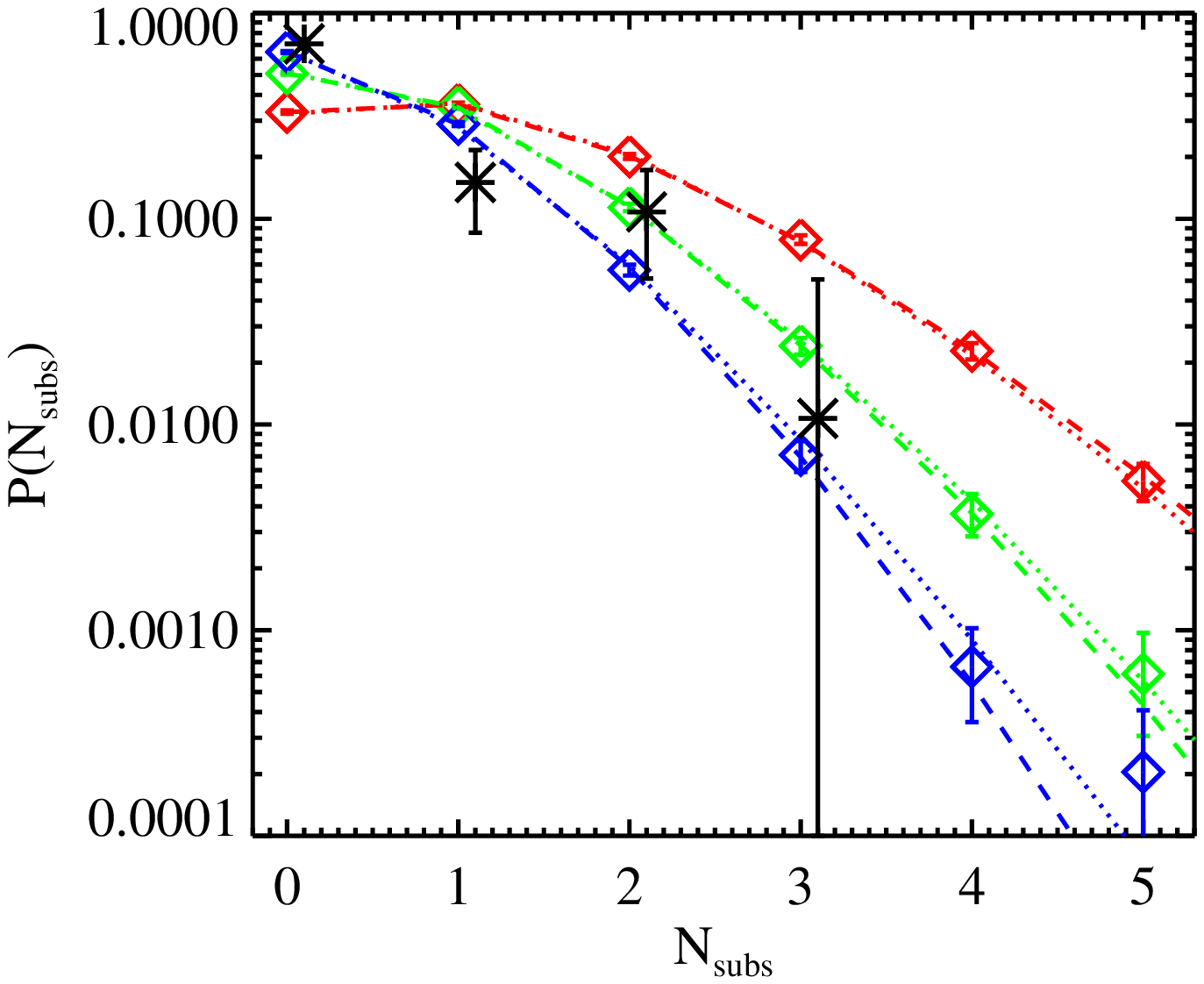} \caption{The
    probability distribution for the abundance of MC-like satellites
    around MW-mass hosts in simulations (colored diamonds).  The left
    panel shows the results for hosts in a mass bin of $\mvir = 1.2\pm
    0.3 \times 10^{12}\msol$; the right panel shows results for hosts
    in a mass bin of $\mvir = 2.6\pm 0.4\times 10^{12}\msol$.  In both
    cases, the red, green, and blue distributions correspond to the
    number of satellites with $\vmax > 50$\kms, $\vmax > 60$\kms, and
    $\vmax > 70$\kms, respectively.  Dashed lines show the best-fit
    negative binomial distribution, and dotted lines show the best-fit
    Poisson distribution.  In both panels, the black points show the
    observational result of \cite{Liu10} for the number of satellites
    within 250 kpc around isolated MW-luminous
    galaxies.} \label{fig:vmax_comp} \end{figure*}

The correlation between subhalo abundance and host mass has been well
studied \citep{Zentner03, Diemand04, Gao04, Zentner05, Klypin10}.
These studies have all found that, when scaled in units of $\mu =
M_{sub}/M_{host}$, larger halos contain more subhalos above a given
$\mu$ than do smaller ones.  This is the expected result from the
hierarchical merging picture of CDM, in which larger objects grow
through the merging of smaller objects and therefore form later
\citep{Blumenthal84,Lacey93}.  This later formation has a two-fold
impact: it reduces the amount of time available to tidally strip a
subhalo, and lowers the host concentration, further reducing the
ability of a host to tidally strip its subhalos \citep{W02, Busha07}.
Motivated by this, we present the abundance of LMC and SMC-mass
subhalos for the 27,000 Bolshoi halos with $\mvir = 2.6\pm 0.9 \times
10^{12}\msol$.  The results can be seen in Figure \ref{fig:vmax_comp}.
Here, the left panel shows the results for selecting hosts with $\mvir
= 0.8-1.7\times 10^{12}\msol$, while the right panel shows a mass
range $\mvir = 1.7-3.4\times 10^{12}\msol$.

For the more massive $\mvir \sim 2.6 \times 10^{12} \msol$ halos, the
cumulative probability of hosting two or more 50 \kms{} halos
increases by a factor of 3 compared to the $1.2\times 10^{12} \msol$
objects (from 8\% to 26\%), while jumping by a factor of 6 for
70\kms{} subhalos (from 1\% to nearly 6\%).  We emphasize that this is
not a contradiction with the results of \cite{Busha10c}, which found
that a halo with exactly two MC-like subhalos was likely to have mass
$1.2\times 10^{12} \msol$.  While the fraction of more massive halos
hosting two MC-like subhalos is larger, the steep slope of the mass
function means that there are many more lower-mass halos.
Additionally, aside from just considering $\vmax$, \cite{Busha10c}
selected halos with satellites like the MCs in terms of location and
three-dimensional velocity, properties not considered in the present
analysis.

Again, these results are in excellent agreement with the work of BK10.
While such an agreement is generally expected from pure collisionless
N-body simulations, such agreement is not as trivial as it may appear.
While the present work uses results from the Bolshoi simulation, run
using the adaptive refinement ART code and the BDM halo finder, BK10
used the Millennium-II simulation run with the TreePM code Gadget-2
with substructures identified using the Subfind algorithm
\citep{subfind}.  Additionally, in both cases, we are concerned with
subhalos close to the mass-limit for the simulations.  Thus, the close
agreement between these two works suggests that numerical effects in
the simulations are well understood at this level.

\subsection{Environmental dependence}
\label{sec:env}

Beyond simple host mass dependence, the size of the cosmological
volume modeled by the Bolshoi simulation allows us to perform further
detailed studies on properties impacting the satellite abundance
distribution.  Here, we quantify the environmental dependence of the
subhalo population by splitting the sample on local density, defined
by \be \delta_{h,r} = {\rho_{h,r} - \bar{\rho_h} \over \bar{\rho_h}} ,
\ee where $\rho_{h,r}$ is the Eulerian density of dark matter
contained in halos larger than $\mvir = 2\times 10^{11}\msol$ (roughly
1000 particles in Bolshoi) defined in a sphere of size $r$.  In the
present analysis, we take $r = 1.0 \hinv\mpc$; we have performed
similar analysis on a range of scales and find this to be the range
where the galaxy halo distribution is the most sensitive to the
presence of massive satellites.

\begin{figure}
\plotone{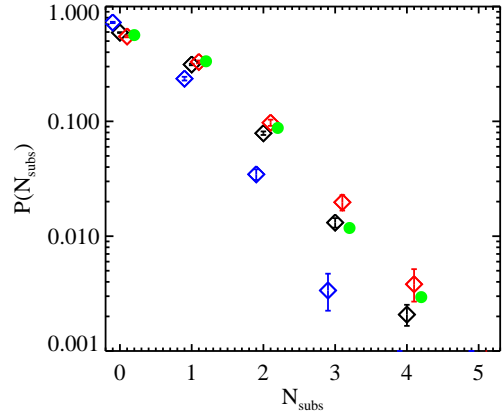}
\caption{The probability that a $1.2 \times 10^{12}\msol$ halo hosts
  \N\ subhalos with $\vmax > 50$\kms, split by local host density.
  The black points represent all halos, while the red and blue points
  represent the most overdense and underdense quartiles, respectively.
  Green circles represent MW-mass halos that were selected to have
  M31-mass neighbors. }
\label{fig:nsats_dens}
\end{figure}

We compare the distribution \PN{} for halos in the top and bottom
quartile distributions to the mean relation in Figure
\ref{fig:nsats_dens}.  There is a clear systematic trend: halos living
in overdense regions (red points) are more likely to host massive
subhalos than those in underdense regions (blue points).  This is
likely due to the fact that halos in denser environments live in more
biased regions, where smaller density perturbations are amplified,
making it easier for smaller halos to form and accrete onto
intermediate mass hosts.  This makes the deviations from the mean
relation in Figure \ref{fig:nsats_dens} a potentially observable
manifestation of assembly bias, which posits that the properties of a
given halo may be determined by properties other than mass, such as
the halo environment.

Selecting for objects in a dense environment increases by almost 25\%
the probability of having $N_{subs} = 2$ or more, from 9\% to just
over 11\%; this effect becomes stronger for larger values of
$N_{subs}$.  The Milky Way is somewhat overdense on this scale, having
a massive companion, M31, as well as a number of surrounding dwarf
galaxies.  Recent estimates put the total mass of the local group at
roughly $5\times 10^{12}\msol$ \citep{Li08}, solidly in a dense
environment on the $\sim$ 1 Mpc scale.  Even if we ignore the
contribution to the local density from all neighboring objects except
M31, the MW lies close to the top quartile of our measured
local density distribution in Bolshoi on a $\sim$ 1 Mpc scale.  We can
begin to directly quantify the impact of such a massive neighbor on
the subhalo population by identifying MW-mass halos which have
M31-like neighbors with $\mvir = (1-3)\times 10^{12}\msol$ at a distance of
700-900 kpc.  These systems are represented by the green circles in
Figure \ref{fig:nsats_dens}, which have a $N_{subs} = 2$ probability
enhanced by 2$\sigma$ relative to the full sample.  Thus, the local
environment around the MW, which is dominated by the presence of M31,
boosts the chance of seeing the Magellanic Clouds in the Milky Way.

\subsection{Probability distributions}
\label{sec:pois}

As an extension of the trends shown in Figure \ref{fig:vmax_comp}, it
is useful to generalize these results to arbitrary host and subhalo
masses by fitting the probability distribution.  A Poisson
distribution seems to be the most natural assumption, and has been
found previously in the literature \citep[e.g.,][]{Kravtsov04a}.
Deviations from a Poisson distribution can be quantified by the
calculation of the second moment, $\alpha_2$, where \be \alpha_2 =
{\langle N(N-1)\rangle \over \langle N \rangle^2} .  \ee For a Poisson
distribution, $\alpha_2 = 1$.  While studies have found some evidence
for a deviation from Poisson, with $\alpha_2 > 1$ \citep{Kravtsov04a,
  Wetzel10}, the distribution has been more fully quantified by BK10.
This work found that the distribution could be much better modeled
using a negative binomial distribution, \be P(N|r,p) = {\Gamma(N+r)
  \over \Gamma(r)\Gamma(N+1)} p^r (1-p)^N
\label{eq:negative_binomial}
\ee
where
\be
p = {1 \over 1 + (\alpha_2^2-1)\langle N \rangle}, \quad r = {1 \over \alpha_2^2 - 1},
\label{eq:p_r}
\ee although they note that a Poisson distribution is a reasonable fit
when $\langle N \rangle$ is low, such as the case of the MCs.  We plot
fits to both a Poisson and a negative binomial distribution to our
measurements as the dotted and dashed lines in Figure
\ref{fig:vmax_comp}.  We again see that, because $\langle N \rangle <
1$, the Poisson distribution provides a reasonable fit, while the
negative binomial distribution fit is excellent (at the expense of
adding an additional parameter).\footnote{Note that, when calculating
  these fits, it is possible to either calculate $\langle N \rangle$
  and $\alpha$ directly from the data, or to treat them as free
  parameters to the fitting algorithm.  However, in practice the
  differences in resulting parameters are less than 5\%.}  Our results
confirm those of BK10, in that the distribution becomes significantly
more non-Poisson for larger values of $\langle N \rangle$.

\begin{figure}
\plotone{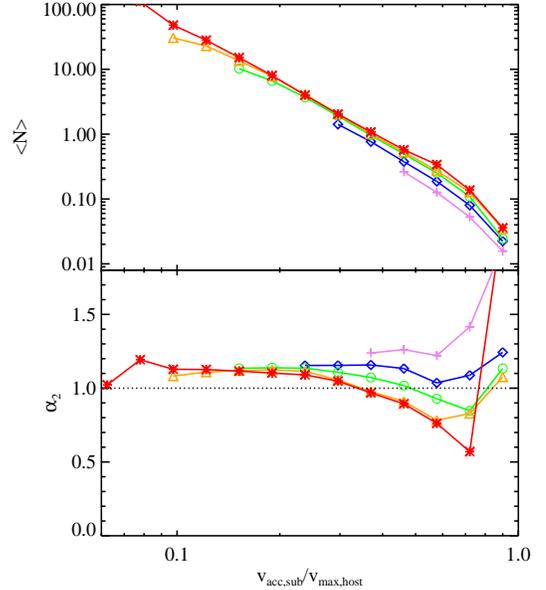}
\caption{{\it Top Panel:} The trend of $\langle N \rangle$ for the
  $v_{acc}$ distribution function for hosts of varying masses.  This
  is equivalent to the subhalo mass function.  Magenta pluses, blue
  diamonds, green circles, and red triangles represent hosts with
  $\vmax = $ 200, 300, 600, and 1000 \kms, respectively.  {\it Bottom
    Panel:} The trend of the scatter in the distribution of the number
  of subhalos above a mass threshold.  The dashed line represents
  $\alpha_2 = 1$, where the negative binomial distribution reduces to
  Poisson.  }
\label{fig:distribution_fits}
\end{figure}

As a final comparison of the Bolshoi halos to previous work in the
literature, we plot in Figure \ref{fig:distribution_fits} the trends
of $\langle N \rangle$ (top panel, similar information to Figure 15 in
\citealt{Klypin10}) and $\alpha_2$ (bottom panel) with mass ratio
$v_{acc,sub}/v_{max,host}$ for a range of host masses.  Here,
$v_{acc,sub}$ is the maximum circular velocity, $v_{max}$, that the
subhalo had at the epoch of accretion.  As seen in the top panel, the
value for $\langle N \rangle$ scales almost self-similarly, but with a
slight increase in overall normalization as has been previously noted
in the literature \citep{Gao04, Klypin10, Gao10}, indicating that more
massive halos contain somewhat more subhalos, proportionally, than
lower mass halos.

More interesting is the trend in $\alpha_2$ with mass, shown in the
lower panel of Figure \ref{fig:distribution_fits}.  Here we see
evidence that, for subhalos with $v_{acc,sub}/v_{max,host} \gsim
0.25$, self-similarity breaks down much more significantly, where
lower-mass hosts tend to have systematically higher values of
$\alpha_2$, which implies increased scatter beyond Poisson statistics.
There are a number of possible explanations for this trend; it may be due
to the fact that lower mass objects exist in a wider range of
environments than higher mass halos.  As shown earlier in Figure
\ref{fig:nsats_dens}, the immediate environment surrounding a halo can
have a significant impact on its subhalo population.  High-mass halos
always form in regions of high bias, meaning that the environmental
impact on a high mass object's subhalo population is minimized.  Low-mass
objects are able to form in both biased and unbiased regions and in
regions near to higher mass halos, creating a larger scatter in their
subhalo populations.  While not shown, we have made analogous plots to
Figure \ref{fig:nsats_dens} for hosts of higher mass and confirm that
these objects typically display weaker environmental dependence in their
subhalo populations. 

It is also interesting to note that previous studies, in particular
\cite{Kravtsov04a} and BK10, have not seen such a trend.  There are a
number of possible explanations for this, which include the improved
$1 \hkpc$ force resolution and $(250 \hMpc)^3$ volume of the Bolshoi
simulation, which allows for both better statistics and more accurate
tracking of subhalos as they orbit around their hosts.  The increased
volume of Bolshoi is particularly important if the trend is driven by
halo environment, since a larger volume is necessary to probe a full
range of environments.  Hints of this trend, however, can be noticed
in Figure 9 of BK10.  While not shown, nearly identical trends in
$\langle N \rangle$ and $\alpha_2$ persist if we consider the mass
ratio $v_{max,sub}/v_{max,host}$, using the maximum circular velocity
of the subhalo at $z = 0$ instead of at the time of accretion.

To summarize, in this section, we have shown that satellites
comparable in size to the LMC and SMC are relatively rare in MW mass
objects, occurring in roughly 9\% of the potential hosts.  Their
abundance is dependent on a number of host properties, such as the
host mass and environment.  The latter can provide a boost at the
$25\%$ level.  The proximity of M31 puts the Milky Way in a higher
density environment, which increases the likelihood of an LMC/SMC pair
to 12\%.  Finally, it was shown that for massive subhalos, the scatter
in the expected number of subhalos systematically decreases with
increasing host mass.  This may be related to the environmental
variations previously observed; lower mass halos are able to form in a
wider range of environments than higher mass objects.  In the next
section, we turn to the task of comparing our simulation results with
observations from the SDSS, which requires the adoption of an
algorithm for adding magnitudes to simulated galaxies.

\section{Satellite Statistics for Galaxies}
\label{sec:luminosities}

We turn next to a detailed comparison of these results with
observational measurements.  The classical satellite galaxies around
the MW have been well-studied, with the SMC and LMC being the only
satellites brighter than $M_V = -16$.  Indeed, the next brightest
object, Sagittarius, with $M_V = -13.1$, is roughly 3 magnitudes
dimmer than the SMC.  Motivated by this, we search for the
distribution of the number of satellites of $M_V \le -16$ in other
galaxies, both simulated and real.  In order to distinguish the
results in this section from the previous sections, we adopt the
formalism where \NS\ refers to the number of satellite galaxies around
a host, identified based on their luminosities.  This contrasts with
mass-selected dark matter subhalos, which we characterized in the
previous sections using the nomenclature \N.

\subsection{The Observational Sample}

Our recent companion paper \citep[hereafter, L10]{Liu10} studied the
population of such LMC and SMC-luminosity objects in the SDSS.  Here,
MW analogs were identified in the SDSS main sample by locating objects
with absolute magnitudes $M_r = -21.2 \pm 0.2$ with no brighter
companions within a line-of-sight cylinder of radius 500 kpc and
length 28 Mpc (corresponding to a redshift of $\pm
1000$~\kms).\footnote{The value for the $r$-band absolute magnitudes
  was determined using the K-correction code of \cite{Blanton07} on a
  sample population of 500,000 SDSS galaxies.}  The abundance of
satellite galaxies was then measured by calculating the likelihood of
having \NS\ neighbors in the photometric sample with apparent
magnitudes 2-4 dimmer in a fixed 150 kpc aperture.  While we discuss
details of their method below, our goal here is to compare our
simulated results to their observational constraints.

\subsection{Modeling Galaxy Luminosities}
\label{sec:sham}

Because internal kinematics are difficult to measure observationally
for dwarfs outside the Local Group, the observations of L10 cannot be
directly compared to the results from the previous section, which
determined abundances based on the $v_{max}$ of the satellites.
Instead, we employ a SubHalo Abundance Matching (SHAM) algorithm to
our simulation to map magnitudes onto dark matter halos
\citep{Kravtsov04a, Conroy06, Wetzel10, Klypin10, Behroozi10}.  Here,
luminosities are assigned by matching the number density of observed
objects brighter than a given luminosity with the number density of
simulated objects greater than a given $\vmax$.  For subhalos, $\vmax$
at the time of accretion is used instead of the present $\vmax(z=0)$.
We also impose a log-normal scatter between halo mass and luminosity.
In our base model, we assume a scatter of 0.16 dex, consistent with
constraints on larger group and cluster mass objects \citep{More09,
  Behroozi10}.

This algorithm has been shown to be extremely successful in matching
several statistical properties of the galaxy distribution, including
the luminosity dependence and redshift scaling of the two-point galaxy
correlation function for galaxies brighter than $M_r < -19$
\citep{Conroy06}, the galaxy-mass correlation function
\citep{Tasitsiomi04}, and galaxy three point statistics
\citep{Marin08}.  \cite{Klypin10} have shown that a similar procedure
can simultaneously match the luminosity function and the Tully-Fisher
relation locally.

\begin{figure}
  \plotone{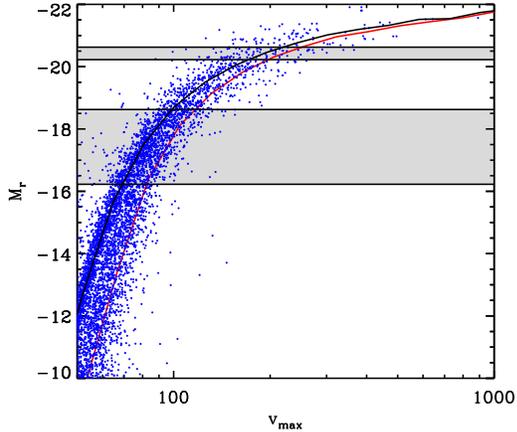}
  \caption{The relation between magnitude and $v_{max}$ for simulated
    halos+SHAM.  The red line represents the relation between $M_r$
    and $v_{acc}$, the maximum circular velocity at the time of
    accretion, which was used to select the magnitude.  The black
    curve and blue points show the relation the between magnitude and
    $\vmax$ at $z=0$ (as opposed to $v_{acc}$).  The upper grey band
    shows our magnitude-criteria for selecting Milky Way-like hosts,
    while the lower gray band represents our range for selecting
    Magellanic Cloud-like satellites. }
\label{fig:mr_vpeak}
\end{figure}

Here, we match halos directly to the binned $r$-band luminosity
function in the local Universe as measured by \cite{Blanton05} for
galaxies with $M_r < -12.8$.  While there is some uncertainty in these
measurements depending on the detailed surface brightness correction,
the objects we are interested in here are bright enough that this
effect is unimportant.  This luminosity function also reproduces the
more recent number densities measured in SDSS DR7 \citep{Zehavi10}
down to $M_r \sim -17$.  The results of this procedure are shown in
Figure \ref{fig:mr_vpeak}.  Here, the red line shows the relation
between $v_{max,acc}$ and magnitude, while the blue points show
$\vmax$ at $z = 0$.  The systematic movement to lower $\vmax$ for many
objects is due to tidal stripping of subhalos.  The resulting
simulated catalog has a distribution of galaxies that is complete down
to $M_r = -15.3$.  \mike{Note that our pure-abundance matching measurements from Figure \ref{fig:mr_vpeak} gives most likely mass estimates for the LMC and SMC to be $\vmax = 85\pm 6$ and $65\pm 6$\kms, respectively.  However, as shown in \cite{TrujilloGomez10}, ignoring the effects of baryons likely increases the maximum circular velocity by roughly 10\% (see their Figure 5), bringing our SHAM mass estimates into excellent agreement with the observations    }

\begin{figure}
  \plotone{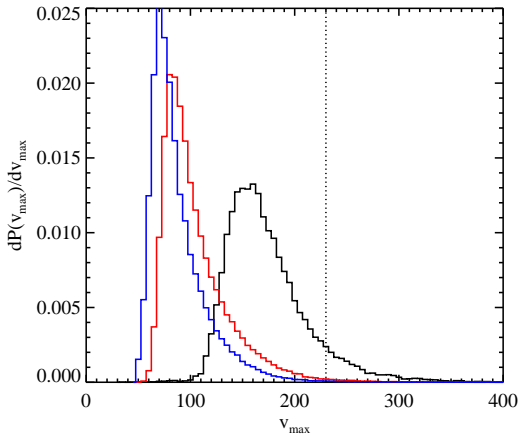}
  \caption{The distribution of $v_{max}$ for simulated galaxies, as
    selected by luminosity.  The black histogram represents galaxies
    that pass our Milky Way selection, with $M_r = -21.2 \pm 0.2$,
    while the blue and red histograms represent $v_{max}$ and
    $v_{acc}$ for our Magellanic Cloud-like subhalos that are 2-4
    magnitudes dimmer than their hosts. The dotted line shows the
    preferred value for $v_{vir}$ of the MW.}
\label{fig:sham_vmax}
\end{figure}

Using luminosities assigned in this way, we now make measurements
analogous to those of L10.  We select for MW candidates in the same
way as L10.  First, we make a mock lightcone by taking the $z=0$ SHAM
catalog of the Bolshoi simulation, placing an observer at the center
of the box, and adding redshift space distortions.  We passively
evolve galaxy magnitudes to the appropriate redshift using the SDSS
constraint $M_r(z) = M_r(z=0.1) + Q(0.1 - 1)$, with $Q = -1.3$
\citep{Blanton03}.  We then identify all isolated objects by selecting
halos with $M_r = -21.2 \pm 0.2$ (the upper grey band in Figure
\ref{fig:mr_vpeak} and excluding objects having a brighter neighbor in
a cylinder of radius 500 kpc and length 28 Mpc.  The resulting $\vmax$
distribution is shown in Figure \ref{fig:sham_vmax}.  This
distribution is somewhat lower than the best current estimates for the
maximum rotational velocity of the MW, ${\rm v}_{rot} = 220$ \kms
\citep{Xue08, Gnedin10}.  This difference should not be surprising
since $\vmax$ includes only the dark matter, while ${\rm v}_{rot}$ is
the total rotational speed of an actual halo that includes the full
impact of baryons.  Further, \cite{TrujilloGomez10} have shown that,
because the Milky Way sits near the peak of the star formation
efficiency, baryons play a significant role in determining the
circular velocity profile of MW mass halos, resulting in a value of
$\vmax$ that is underestimated in dark matter-only simulations.
However, the virial mass of the galaxies that pass our MW selection
peaks at $1.3 \times 10^{12} \msol$, in excellent agreement with the
estimated MW mass from \cite{Busha10c}.

We select for MC-analogs using a method similar to L10, identifying
all satellites within the virial radius of our MW analogs that are 2-4
magnitudes dimmer (lower grey band in Figure \ref{fig:mr_vpeak}).  The
most likely $\vmax$ values for our MC analogs are $\sim$ 80 and $\sim$
70 \kms{}, in agreement with the 1-$\sigma$ constraints from
observations\citep{VanDerMarel02, Stanimirovic04, Harris06}.

\subsection{The Definitions of a Satellite}

Before presenting detailed comparisons between the L10 observational
data and our model, we note that the definition of a ``satellite''
differs between the two analyses.  Because L10 used a photometric
sample with hosts selected on luminosity and no spectroscopy for their
satellites, the number of satellites was defined as the number of
photometric pairs around a host. Specifically, they identified
isolated MW-luminosity galaxies in the SDSS spectroscopic sample and
counted the number of photometric pairs with $\Delta m_r = 2-4$ within
a fixed angular separation, nominally taken to be 150 kpc.  They then
applied a background subtraction to account for projection effects and
correlated structures.  While our SHAM catalog will let us select
isolated host objects in an analogous way, because modeling the full
distribution of galaxy apparrent magnitudes is beyond the scope of our
model, we do not perform an identical calculation.  Instead, we can
select an appropriate definition of a satellite galaxy to mimic the
observational effects.

L10 presented two methods for background subtraction, which were used
to calculate the number of satellites around a host galaxy.  The
first, which we refer to as isotropic subtraction, assumed that
background objects had, on average, no dependence on position
(implicitly ignoring the impact of correlated structures), to
calculate an average background for their entire host population.
Using this assumption, it is possible to calculate the average
background to very high statistical accuracy.  Their second method,
which we refer to as annulus subtraction, measured the background
object counts around each galaxy by considering the region just
outside the projected region surrounding the galaxy.  For
completeness, we make comparisons to both methods.

Because it removes the effects of correlated structures, comparison to
the annulus subtraction method of L10 is straightforward.  We define
as a satellite galaxy any object within the appropriate spherical
volume (taken to be 150 kpc in L10) that has an absolute magnitude
that is 2-4 mag dimmer than its host.  To compare to the isotropic
subtraction method, which neglects the impact of correlated structures
along the line of sight (see Section 3.2 of L10 for further
discussion) requires additional modeling.  We account for this by
defining a satellite to be any object inside a cylinder of radius 150
kpc and length $r_0$, where $r_0$ is the correlation scale for the
Milky Way host-satellite analog.

\begin{figure}
  \plotone{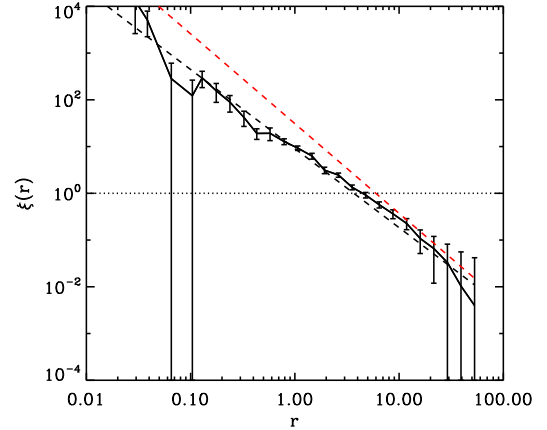} \caption{The 3D cross-correlation
    function, $\xi(r)_{\rm MW-MC}$, between Milky Way like hosts and
    all LMC and SMC magnitude satellite galaxies.  The black dashed
    line represents a power-law fit to $\xi(r)$.  For reference, the
    dotted red line denotes the auto-correlation between galaxies with
    $-21 < M_r -5\log(h)= -20$ in SDSS DR7 as measured in
    \cite{Zehavi10}.  } \label{fig:CrossCorrelation} \end{figure}

This cross-correlation function between MW-luminous galaxies and
objects 2-4 magnitudes dimmer is plotted in figure
\ref{fig:CrossCorrelation}, and is well fit by a power law, \be \xi(r)
= (r/r_0)^{-\gamma} \ee with $r_0$ = 5.3 Mpc and $\gamma = 1.7$, with
a similar correlation length to measurements for all galaxies of
MW-like magnitudes inferred from \cite{Zehavi10}.  These authors
measured $r_0$ = 5.98 Mpc/$h$ and $\gamma$ = 1.9 for the
autocorrelation function of galaxies with $M_r = -20.7$ to $-21.7$.
We therefore consider a cylinder with radius 150 kpc and length 5.3
Mpc (the correlation length) to be the appropriate aperture for
satellite selection to compare to the isotropic background subtraction
of L10.

Coincidentally, this correlation scale is very similar to the size of
the expected redshift-space distortions.  \cite{Evrard08} showed that
a $\sim 10^{12} \msol$ halo has a typical velocity dispersion of 125
\kms; at $z = 0.08$, where the bulk of our simulated halos live, this
corresponds to a 3.7 Mpc spread due to the redshifts of the satellite
population.  A cylinder of this length thus represents the most
accurate measurements one could make for the number of satellites
around a host using spectroscopic data (without including some sort of
additional background subtraction).

\begin{figure}
  \plotone{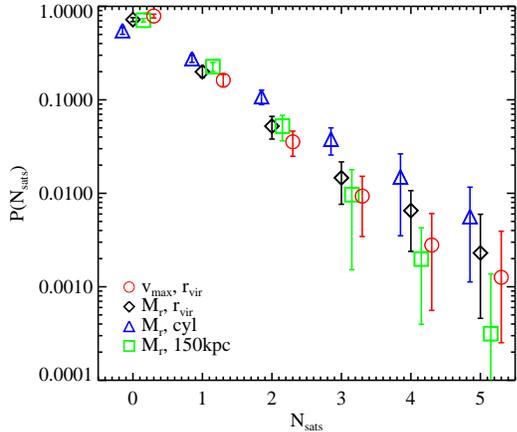}
 \caption{The probability of having \NS{} LMC and SMC-type objects in
    simulations based on SHAM and $v_{max}$ for various satellite
    definitions.  Red circles count objects with $v_{max} > 50$ \kms
    within the virial radius; the black diamonds, blue triangles, and
    green boxes count objects 2-4 magnitudes dimmer than their host
    using a SHAM model with 0.16 dex scatter.  The black diamonds were
    calculated using all objects inside the virial radius.  The blue
    triangles and green boxes are equivalent to the isotropic and
    annulus background subtractions used in L10: a cylindrical
    aperture of radius 150 kpc and length 5.3 Mpc (blue triangles),
    and a fixed 150 kpc spherical aperture (green boxes).}
\label{fig:nsats_comp} \end{figure}

In Figure \ref{fig:nsats_comp} we compare the \PN\ distribution for
four different satellite definitions: spherical and cylindrical
apertures (which correspond to the L10 results using their annulus and
isotropic background subtractions methods, respectively) as well as
selecting satellites based on $\vmax$ and abundance-matched magnitude.
Here, the red circles consider subhalos within the virial radius of
their host that have $v_{max} = 50-80 $\kms.  The black diamonds, blue
triangles, and green boxes all use magnitude-selected satellites,
counting objects 2-4 magnitudes dimmer than their host using the
aforementioned SHAM model with cuts analogous to the observational
criterion of L10.  The red diamonds represent the ``correct'' number
of satellites, using all objects inside the virial radius, The blue
triangles use a cylindrical aperture, analogous to the isotropic
background subtraction of L10, while the green boxes use a spherical
aperture, equivalent to the annulus subtraction method of L10.  All
error bars were calculated using the jackknife method with 10
sub-regions.

A number of trends are readily apparent for the relations in Figure
\ref{fig:nsats_comp}.  First, there is generally good agreement
between the \PN\ distribution for objects based on $\vmax$ (red
circles) with our SHAM magnitudes (black diamonds).  This is to be
expected if SHAM preserves the observed mass-luminosity relation in
this regime, something that will only happen if an appropriate mass
function is used.  It is also interesting to note that our
calculations for the cylindrical aperture (blue triangles) always lie
above the other measurements for $N_{subs} \ge 1$.  This shows that
correlated structures are an important effect which is typically
comparable to the signal from satellites that are formally inside the
halo's virial radius.  Our measurements using a fixed 0.15 Mpc
cylinder (green box) lie below other measurements for higher values of
\NS.  This should be unsurprising, since the typical virial radius for
our selected halos is roughly 250 kpc.  It is interesting to note,
however, that the cumulative probability for hosting (at least) one
satellite differs by less than 5\% when calculated using the virial
radius instead of a radius of 150 kpc, while this difference increases
rapidly for higher values of \NS.  All of these trends indicate the
importance of careful background subtraction to eliminate this signal
(such as the annulus subtraction of L10), or otherwise of comparing
directly to simulations with similar selection criteria so as to
properly interpret results.

As concerns the length of our cylindrical aperture, we obtain almost
identical results for any length beyond $\sim 1$Mpc, beyond which the
correlation function falls off substantially.  Varying the cylinder
length from $2-8$ Mpc changes our results by roughly five percent,
comparable to the size of our statistical errors.  Thus, the
comparison between satellites defined using our cylindrical aperture
and the isotropic background subtraction technique of L10 should be
robust.

The numerical values for many of these probabilities --- selection by
$r_{vir}$, as well as our cylindrical and fixed spherical apertures ---
are presented in Table \ref{table:probabilities}.  Note that, while we
are only counting objects as satellites if they are 2-4 magnitudes
dimmer than their hosts for the most accurate comparison with L10,
these numbers are essentially identical to considering a threshold
sample of objects 0-4 magnitudes dimmer than their (MW-like) hosts.
Switching to this definition creates only percent-level changes, well
within the statistical error bars.

\subsection{Comparisons Between Simulations and Observations}
\label{sec:comparisons}
\begin{figure}
 \plotone{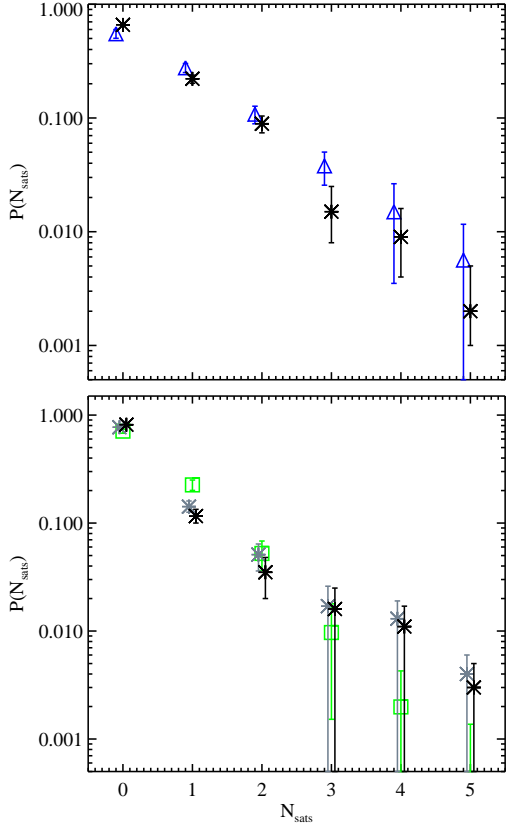} 
  \caption{Comparisons between the \PNS \ distribution of SMC-luminous
    satellites around MW-luminous hosts in the Bolshoi simulation and
    the observational results of L10.  {\it Top Panel:} Comparison of
    the isotropic background subtraction model from L10.  Black
    asterisks represent the abundances from SDSS hosts.  The blue
    triangles denote the abundance of objects inside an aperture that
    most closely corresponds to the isotropic subtraction model: a
    fixed cylindrical aperture with radius 150 kpc and length 5.2
    Mpc. {\it Lower Panel:} Same as the upper panel, but now comparing
    the annulus subtraction model from L10 (black and gray asterisks)
    to the simulation-equivalent of a fixed 150 kpc spherical aperture
    for identifying satellite galaxies around their hosts (green
    squares).  For the L10 data, the black points represent our
    fiducial measurements, while the gray points add a systematic
    uncertainty to the background subtraction.}
\label{fig:nsats_obs_sim} 
\end{figure}

Figure \ref{fig:nsats_obs_sim} and Table \ref{table:probabilities}
show the main result of our analysis, making direct comparison with
the results of L10.  In the top panel, we compare the L10 \PNS
~measurements using their isotropic background subtraction (black
asterisks) with our SHAM results using subhalos inside a fixed
cylinder with radius 150 kpc and length 5.3 Mpc (blue triangles;
cylinder column in Table 2) that closely mimics their selection
criteria.  The bottom panel shows the comparison between the L10
annulus subtraction (black and gray asterisks) and our measurement of
\NS\ within a 150 kpc spherical aperture (green squares; Sphere column
in Table 2).  The black and gray asterisks in the lower panel
represent the range of measurements based on uncertainty in selecting
an optimal background subtraction.  The difference between these sets
of points thus gives an estimate for the size of the systematic errors
in the measurements.  Note that, after making their background
subtraction, L10 also calculate a systematic adjustment due to
catastrophic photo-z failures (see section 4.2 of L10).  While we have
included those corrections in Figure \ref{fig:nsats_obs_sim}, we have
explicitly kept the lower error bar consistent with zero for any
measurement that was consistent with zero at the 2-$\sigma$ level
before this systematic correction was included.  For reference, Table
\ref{table:probabilities} also gives the \PNS\ distribution for proper
subhalos, i.e., objects within $r_{vir}$ that are 2-4 magnitudes
dimmer than their host.

The agreement between simulations and observations in Figure
\ref{fig:nsats_obs_sim} is generally very good.  We take this to
indicate a success for both the \LCDM and SHAM models, particularly
indicating, with excellent statistics, that the HOD predicted by
Bolshoi for Milky Way-like objects at the massive end matches
observations, and there is no evidence for either a missing or excess
satellite problem for these objects.  Considering our comparison
between the cylindrical/isotropic sample, at no \NS\ do the two
results differ by more than $\sim 2.5 \sigma$, and in most situations
agree to better than 1 $\sigma$.  The comparison fares slightly worse
for the spherical/annulus sample.  Here, the simulations underpredicts
the number of $N_{sats} = 0$ systems, while over-predicting the
$N_{sats} = 1$ systems at the 3 $\sigma$ level.  While this could
indicate a real disagreement of the simulation and galaxy formation
model with the data at this mass scale, there are a number of other
possibilities.  First, it is possible that the annulus-subtraction
method of L10 over-corrects the background subtraction.  This has been
tested in simulations, using their algorithm to estimate the number of
objects within 150kpc around our halos centers using our cylindrical
counts.  The result, is an over preduction of the \NS\ = 0 term by 6\%
and an underprediction of the \NS\ = 1 term by 5\%, larger than the
estimated systematical uncertainties in Table 1 of L10 (the changes in
the higher \NS\ terms are of similar size to their estimated
systematic uncertainty, and are therefore well characterized by the
gray points in Figure \ref{fig:nsats_obs_sim}).  Such corrections
would bring the these terms within 2- and 3- $\sigma$ of our simulated
estimates.  It is also important to note that the predictions from
simulations depend on the assumed scatter between mass and luminosity.
This will be further discussed below.

\begin{table} 
  \caption{Probability of hosting \NS\ satellite galaxies brighter than $M_{r,host} + 4$ around a Milky Way magnitude host halo within various apertures for both our simulated results and the observational measurements from L10.}  
\begin{center} 
\begin{tabular}{| c || c | c c | c c|} \hline
      & $r_{vir}^{a}$ & Cylinder$^b$ & Isotropic$^c$ & Sphere$^d$ & Annulus$^c$\\
      \hline \hline
     $P(0)$ & $55\pm 4\%$ & $55 \pm 5\%$ & $66^{+1.6}_{-1.3}$ & $71 \pm 3\%$ & $81^{+1.5}_{-1.4}$\\
      $P(1)$ & $28\pm 3\%$ & $28 \pm 3\%$ & $22^{+1.7}_{-1.9}$ & $23 \pm 2\%$ & $12^{+1.8}_{-1.6}$\\
      $P(2)$ & $11\pm 1.4\%$ & $11\pm 2\%$ & $8.9^{+1.5}_{-1.5}$ & $5.2 \pm 1.6\%$ & $3.5^{+1.3}_{-1.5}$\\
      $P(3)$ & $3.8\pm 0.7\%$ & $3.8\pm 1.2\%$ & $1.5^{+1.0}_{-0.7}$ & $1.0\pm 0.8\%$ & $1.6^{+0.9}_{-1.6}$\\
      $P(4)$ & $1.55\pm 0.4\%$ & $1.5 \pm 1.1\%$ & $0.9^{+0.7}_{0.5}$ & $0.2 \pm 0.1\%$ & $1.1^{+0.6}_{-1.1}$\\
      \hline \end{tabular} 
\end{center}
$^{a}$halos contained within the virial radius of the host\\
$^b$halos within a cylinder with radius 150 kpc and length 5.2 Mpc\\
$^c$L10 results counting objects within a projected 150 kpc and an isotropic background subtraction\\
$^d$halos within a sphere of radius150 kpc\\
$^e$L10 results counting objects within a projected 150 kpc using an annulus background subtraction\\

\label{table:probabilities}
\end{table}

\begin{figure}
  \plotone{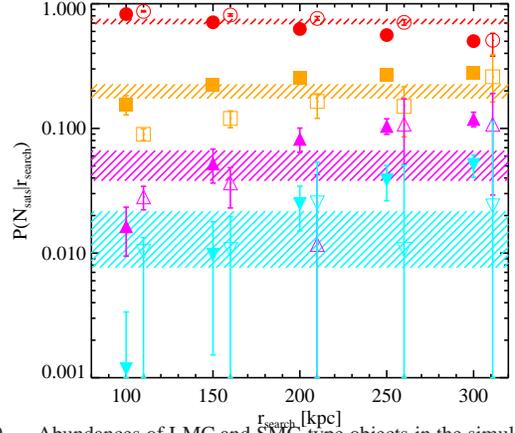}
  \caption{Abundances of LMC and SMC-type objects in the simulation,
    as a function of search radius.  In all cases the filled symbols
    were calculated using Bolshoi+SHAM, while the open symbols
    represent SDSS measurements from L10.  Here, the red circles,
    orange squares, magenta triangles, and cyan inverse-triangles
    represent the probability for a Milky Way like galaxy to have 0,
    1, 2, or 3 subhalos, respectively.  For the Bolshoi objects, the
    probabilities were calculated using a spherical aperture with
    varying radius.  The SDSS points used the annulus subtraction
    method of L10.  The hatched lines represent the probabilities for
    hosting \NS subhalos inside the virial radius of the host.}
\label{fig:nsats_rad}
\end{figure}

Looking at the SDSS/Bolshoi comparisons in more detail, we can further
investigate the dependence of the \PNS \ distribution on the aperture
size, as done in L10.  For this, we determine the probability of
finding \NS\ satellite galaxies in a spherical aperture around their
host, where the radius is varied from 100 to 250 kpc using the annulus
background subtraction.  The results are shown in Figure
\ref{fig:nsats_rad}.  Here, the red circles, orange squares, magenta
triangles, and cyan inverse-triangles represent the probability of
having 0, 1, 2, or 3 subhalos.  Open symbols represent SDSS
measurements, while filled symbols come from our Bolshoi+SHAM model,
while the hatched lines represent the \PNS\ distribution for
satellites within the virial radius of their hosts.  Again, we see
good agreement between simulations and observations, although we note
the persistence of higher values in the simulation for the \NS\ = 1
measurement.  However, the other values of \NS\ are in much better
agreement.

\begin{figure}
  \plotone{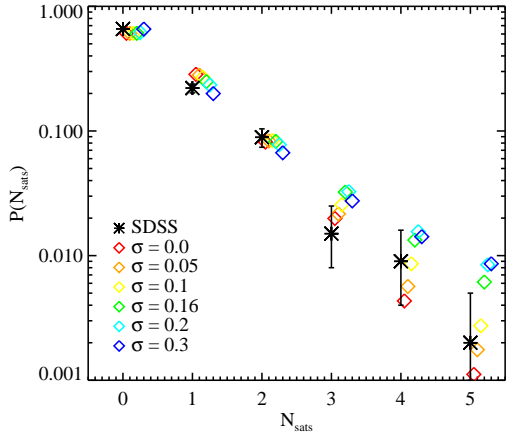} \caption{Dependence of satellite
    probabilities on scatter in the relationship between halo mass and
    galaxy luminosity. Here, we compare the \PNS\ distribution
    calculted using our cylindrical method to the L10
    measurements using an isotropic background subtraction.  
    In each case, the model assumes a constant 
    value of $\sigma$ corresponding to the scatter in log L at a given
    $\vacc$.  Larger scatter values tend to have more systems with
    many satellites, largely because more high mass hosts are
    scattered into the sample. }
\label{fig:nsats_scatter}
\end{figure}

One of the appeals of the SHAM algorithm is that it has so few free
parameters, in this particular case just the scatter (although there
are additional implicit assumptions, e.g., about subhalo stripping and
star formation after satellite accretion).  Figure
\ref{fig:nsats_scatter} shows how the scatter impacts this result.
Here, we present the probability distribution for an isolated host
with $M_r = -21.2$ to host \NS\ satellites brighter than $-16.5$ as we
vary the scatter in the SHAM model from 0 to 0.3 dex using our
cylindrical method for selecting satellites.  While none of these
models are ruled out by the data (L10 isotropic model), a number of
systematic trends are present.  In particular, there is a systematic
increase in the probability of hosting 3 or more satellites, along
with a decrease in the probability of hosting 1 satellite (by up to
5\%).  This trend is caused not by scatter in the mass--luminosity
relation of the satellite galaxies, but in the mass--luminosity
relation of the host objects.  As scatter increases, higher and higher
mass hosts scatter into our magnitude-selected sample, bringing with
them their richer subhalo populations. As we shall discuss in the next
section, this has the effect of creating a tail to the
negative-binomial distribution that sets the probability for a host to
have \NS\ satellites.  

It is also worth noting that the dark blue points representing a
scatter of $\sigma = 0.3$ dex look to be violating the trend of an
increasing high-\NS\ probability.  This is not due to any special
behavior in this regime, but more represents the limits of our
simulation resolution.  As scatter increases, objects with $M_r -
5\log h> -16.5$, which nominally have $v_{acc} = 90$ \kms using a
scatter-free SHAM, have infall masses as low as 65 \kms in our $\sigma
= 0.3$ dex model.  \mike{Because tidal stripping can reduce the $\vmax$ of a subhalo by a factor close to 2 \citep{Wetzel10, Guo10}, this puts such objects dangerously close to the Bolshoi completeness limit
of 50 \kms.  It should be noted, however, that the completeness of subhalos in the Bolshoi simulation is highly dependent on the host mass.  Because of stripping of subhalos after accretion, we
expect to be losing potentially 10-15\% of the satellite population around $\sim 10^{12}\hinv\msol$ mass halos
with scatter this high.}  It's important to appreciate the high
resolution necessary to fully model a galaxy population using SHAM.
While a scatter of $\sigma = 0.3$ is likely ruled out for massive
galaxies \citep{Behroozi10}, if such scatter were appropriate the
Bolshoi simulation would only be able to populate galaxies down to
$M_r \sim -17.5$ while losing fewer than 10\% of the objects.

Finally, we should note that there is no compelling reason to assume
that both massive and low-mass galaxies exhibit the same scatter in
the mass-luminosity relation.  We have investigated this by
quantifying the variations in the \PNS\ distribution where we keep the
mass-luminosity scatter of the hosts fixed at 0.16 and vary the
scatter of the subhalos from 0-0.3.  While we see similar trends with,
e.g., the \NS\ = 1 term dropping with increased scatter, the effect is
significantly weaker than shown in Figure \ref{fig:nsats_scatter},
where we allow the scatter in the host mass-luminosity relation to
vary.  While it is very possible that the scatter should be even
higher than 0.3 for our satellite galaxies, the 50 \kms resolution
limit of Bolshoi prevents us from exploring a wider range.  Thus, over
the ranger we are able to probe, the scatter in the host
mass-luminosity relation, which sets the range of host masses sampled,
has a larger impact that the scatter in the satellite mass-luminosity
relation.

\subsection{Environmental Effects}

\begin{figure}
  \plotone{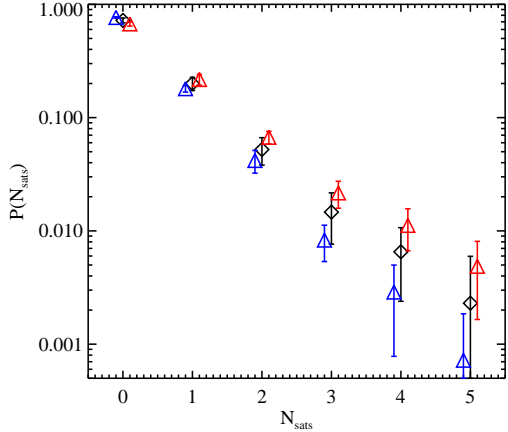} 
  \caption{The probability distribution for a host galaxy to host $NS$
    satellites within its virial radius split by environment.  The black diamonds represent the
    distribution for all isolated hosts (and are identical to the
    black diamonds in Figure \ref{fig:nsats_comp}).  The red and blue
    triangles represent the distribution for hosts in high and low
    density environments.  Here, we use the total luminosity in
    galaxies brighter than $M_r = -20.8$ as our proxy for density.  }
\label{fig:nsats_mag_dens}
\end{figure}

We now revisit the discussion regarding the environmental impact on the
subhalo population of \S \ref{sec:env},  using observationally motivated
environmental measures.  Recall that in \S \ref{sec:env}, we defined
the local environment as the dark matter density in collapsed objects
larger than $2 \times 10^{11} \msol$.  Here, we preform a similar
analysis using luminosity as an environmental proxy.  Specifically,
using the isolated central galaxies from our SHAM catalog as selected
above, we measure the total luminosity from galaxies brighter than
$M_r = -20.8$ (which corresponds to $M_r - 5\log(h) = -20$ in the
usual SDSS units) within a sphere of 1 Mpc around our hosts, which we
take to be our environment proxy.  We then split the sample into our
most and least luminous (i.e., dense) quartiles and re-measure the
\PNS\ distribution for satellites inside their host virial radii.  The
results are shown in Figure \ref{fig:nsats_mag_dens}.  Here, the black
points represent the distribution for all isolated hosts, while the
red and blue represent those in high/low density environments.  As
before, we see a clear offset for galaxies in different environments.
While the differences are only at the 1-2 sigma levels, high-density
halos are systematically more likely to have massive subhalos than
those in low density environments.  In particular, the underdense
halos are 14\% more likely to host {\em no} massive subhalos than
their dense counterparts.  Similarly, halos in dense environments are
twice as likely to host two or more subhalos than those in low density
environments.  This effect should therefore be an observable
manifestation of assembly bias.

\section{Statistics of Satellites in Luminosity-selected Hosts}
\label{sec:generalproperties}

As noted previously, the inclusion of scatter in the halo
$\vmax$-galaxy luminosity SHAM creates a significant tail in the
probability distribution for a host to have \NS\ satellite galaxies.
Consequently, the negative binomial distribution that was found to
describe the $P(N_{\rm subs})$ distribution for the subhalos in a
mass-selected host does not produce a good fit to $P(N_{\rm sats})$
when applied to luminosity-selected galaxies.  Additionally, the
change in star formation efficiency with host halo mass breaks the
self-similarity that was observed for host halos \citep[][and
references therein]{Behroozi10}. In this section, we quantify the
\PNS\ distribution for satellite galaxies around a luminosity-selected
host.

\begin{figure}
  \plotone{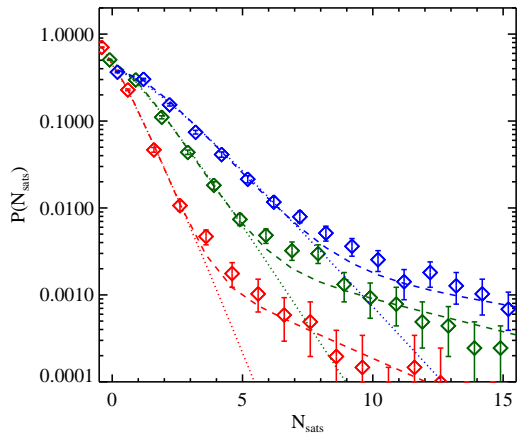} \caption{Probability of hosting
    \NS\ satellite galaxies in a Milky Way-luminosity host, for
    various satellite selection criteria.  The red, green, and blue
    points represent satellites brighter than $M_r = -19.8, -18.2$,
    and $-16.8$.  Dotted lines represent the best fit negative
    binomial distribution, while the dashed line shows the best fit
    negative binomial plus power-law distribution.  }
\label{fig:nsats_with_fit}
\end{figure}

In Figure \ref{fig:nsats_with_fit}, we show a series of probability
distributions for a $M_r = -20.5$ galaxy to host \NS\ subhalos
brighter than -19.8, -18.3 and -16.8 (equivalent to $M_r - 5 \log(h) =
-19, -17.5, -16$).  This plot was generated using a scatter of $\sigma
= 0.16$ dex.  A low probability tail clearly extends to high \NS.
Such a tail was not present when considering $\vmax$-selected hosts
and subhalos in section \ref{sec:pois}.  This is emphasized by the
dotted lines, which present the negative binomial distributions,
equation \ref{eq:negative_binomial}, that characterized the \PN\
distribution.  While the negative binomial provided an excellent fit
to the data for our mass-selected objects, it clearly fails here.
While this tail may not appear to be a serious concern, since it is
not important until \PNS$\sim$ 1\%, large surveys such as the SDSS
contain more than 10,000 massive clusters, so even these small effects
may be visible.

\begin{figure*}
\includegraphics[angle=90,width=6.5in]{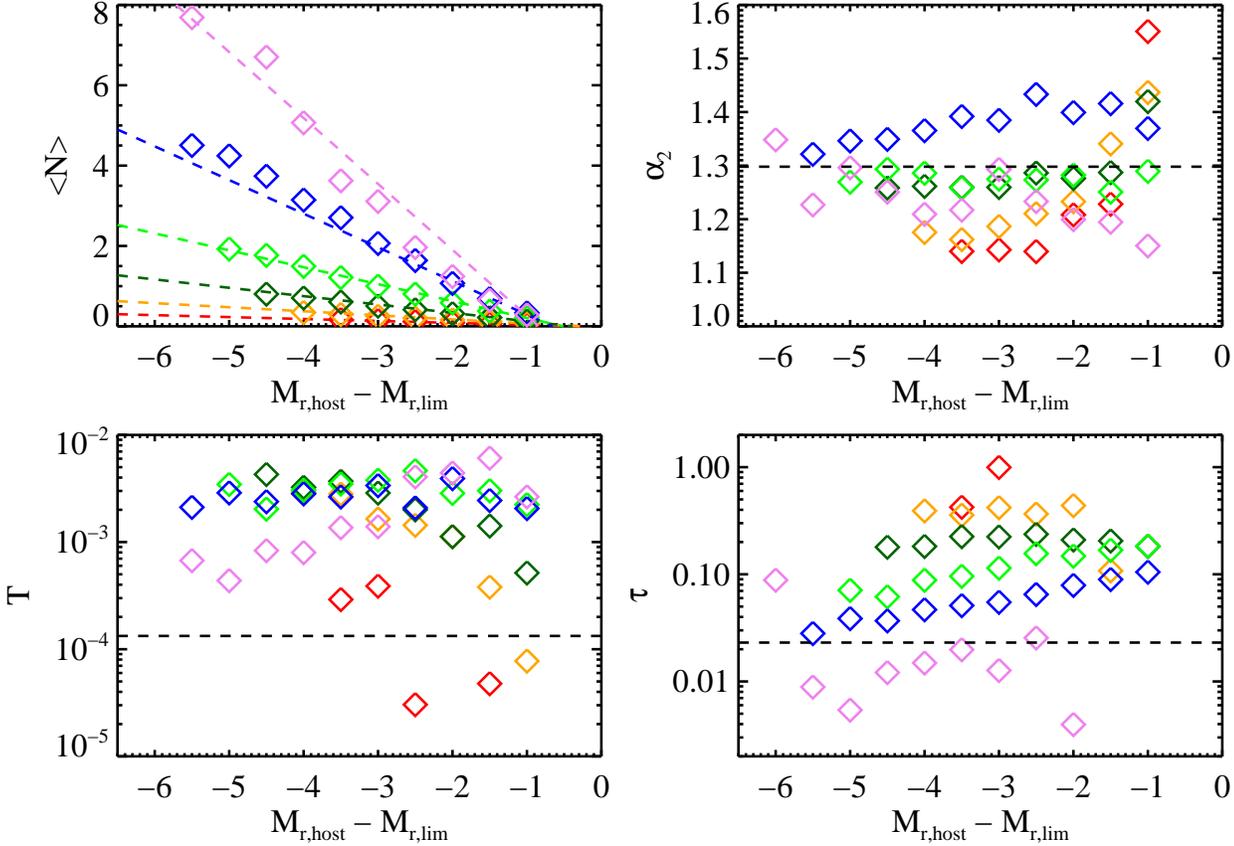}
\caption{The fit parameters specifying the P(N) distribution (equation
  \ref{eq:fit}) as a function limiting satellite magnitude,
  represented as a magnitude difference (luminosity ratio) Different
  colored points represent different host populations: violet, blue,
  green, dark green, orange, and red points are hosts with $M_r =
  -19.8, -20.3, -20.8, -21.3, -21.8$, and $-22.3$.  Dashed lines
  represent out fits to these distributions, as given in Equation
  \ref{eq:full_fit}. }
\label{fig:fit_parameters}
\end{figure*}

As a better fit, we present a modified binomial distribution by adding
a power law tail, 
\be P(N|r,p,T,\tau) = {\Gamma(N+r) \over
  \Gamma(r)\Gamma(N+1)} p^r (1-p)^N + Te^{-\tau N}.
\label{eq:fit}
\ee Here, the first term is just the negative binomial, with $r$ and
$p$ functions of $\alpha_2$ and $\langle N \rangle$, as defined in
Equation \ref{eq:p_r}.  The final term models the high-\NS\ tail that
comes about as a result of the $\vmax$-luminosity scatter, where $T$
and $\tau$ control its slope and amplitude.  The dashed lines in
Figure \ref{fig:nsats_with_fit} show the best fits to Equation
\ref{eq:fit}.  This clearly provides a significantly improved fit to
the simulation measurements, albeit at the expense of adding two new
fit parameters.  Note, however, that the fit tends to have problems in
the cross-over region and that the transition to the high-\NS\ tail is
somewhat more gradual than this model is able to reproduce.  While we
leave this for future investigations, it is worth noting that changing
the scatter in our SHAM model has a direct impact on $T$ and $\tau$.
High values of scatter will have a more pronounced tail, which will
cause both the amplitude and slope of this tail to increase.  For
simplicity, for the rest of this work, we will limit ourselves to
using a model with $\sigma = 0.16$ dex.

While we could present our best-fitting parameters to the
distributions in Figure \ref{fig:nsats_with_fit}, a much more useful
process is to model the \PNS\ distribution as a function of both host
magnitude and limiting satellite magnitude, as has been done for
mass-selected subhalos \citep[see, i.e.,][]{Gao04}.  Unfortunately,
this proves to be much more difficult for galaxies than for halos.
Because the dark matter power spectrum is mostly featureless in the
relevant regime, large halos are nearly self-similar versions of
smaller halos.  As an example of this, the top panel of Figure
\ref{fig:distribution_fits} shows the mean number of satellites above
a given mass cut for mass-selected host halos.  This distribution
exhibits a power-law behavior with nearly identical parameters
regardless of the host mass. This is not the case for
magnitude-selected objects.

Figure \ref{fig:fit_parameters} presents the best-fit parameters for
the satellite \PNS\ distribution as modeled by Equation \ref{eq:fit}
around galaxies as a function of limiting subhalo luminosity (i.e.,
magnitude difference).  The red, orange, yellow, dark green, green,
and blue points represent hosts with $M_r = -20.3, -20.8, -21.3,
-21.8, -22.3$, and $-22.8$.  As can be clearly seen in the top left
panel, which shows the trend of the mean number of satellite galaxies
($\langle N \rangle$) with luminosity ratio, galaxies do not scale
self-similarly.  At a given luminosity ratio, brighter galaxies are
significantly more likely to have a larger \NS\ (expected number of
satellites) than dimmer galaxies.

This is a manifestation of the varying star formation efficiency with
halo mass as seen in, e.g., Figure \ref{fig:mr_vpeak}.  For masses
above $10^{12}$ \msol, star formation efficiency becomes less
efficient, so the rate of increase of $M_r$ with $v_{max}$ is shallow.
At masses lower than $10^{12}$ \msol, star formation efficiency drops
rapidly with decreasing mass, leading to a much steeper rate of change
of $M_r$ with respect to $v_{max}$.  Thus, for a fixed bin in $M_r$,
the corresponding logarithmic range in $v_{max}$ will be much less for
low-mass galaxies as compared to high-mass galaxies.  As such, while
the $\langle N \rangle$ values for mass-selected subhalos lie on a
single line in Figure \ref{fig:distribution_fits}, luminosity-selected
halos do not scale in such a manner, resulting in the different lines
in the top-left panel of Figure \ref{fig:fit_parameters}: brighter
host galaxies have more satellites at a fixed luminosity ratio,
$L_{sat}/L_{host}$, than dimmer hosts.

A number of the trends present in Figure \ref{fig:fit_parameters} can
be readily understood in terms of simple models.  As discussed above,
the behavior of $\langle N \rangle$ is simply a manifestation of the
changing star formation efficiency with halo mass.  Similarly, the
trends in $\tau$ can be understood though simple scaling relations.
First, consider a fixed luminosity ratio, $|M_{r,host} - M_{r,lim}|$.
Since brighter hosts correspond to more massive halos, these objects
live in a steeper part of the mass function, where a fixed scatter
creates a broader selection of host masses, the larger of which are
more likely to host bright satellites.  Because of this, the
exponential tail falls off more slowly, resulting in a lower $\tau$.
The trend of $\tau$ increasing with luminosity ratio for a fixed host
mass.

\begin{figure*}
  \plotone{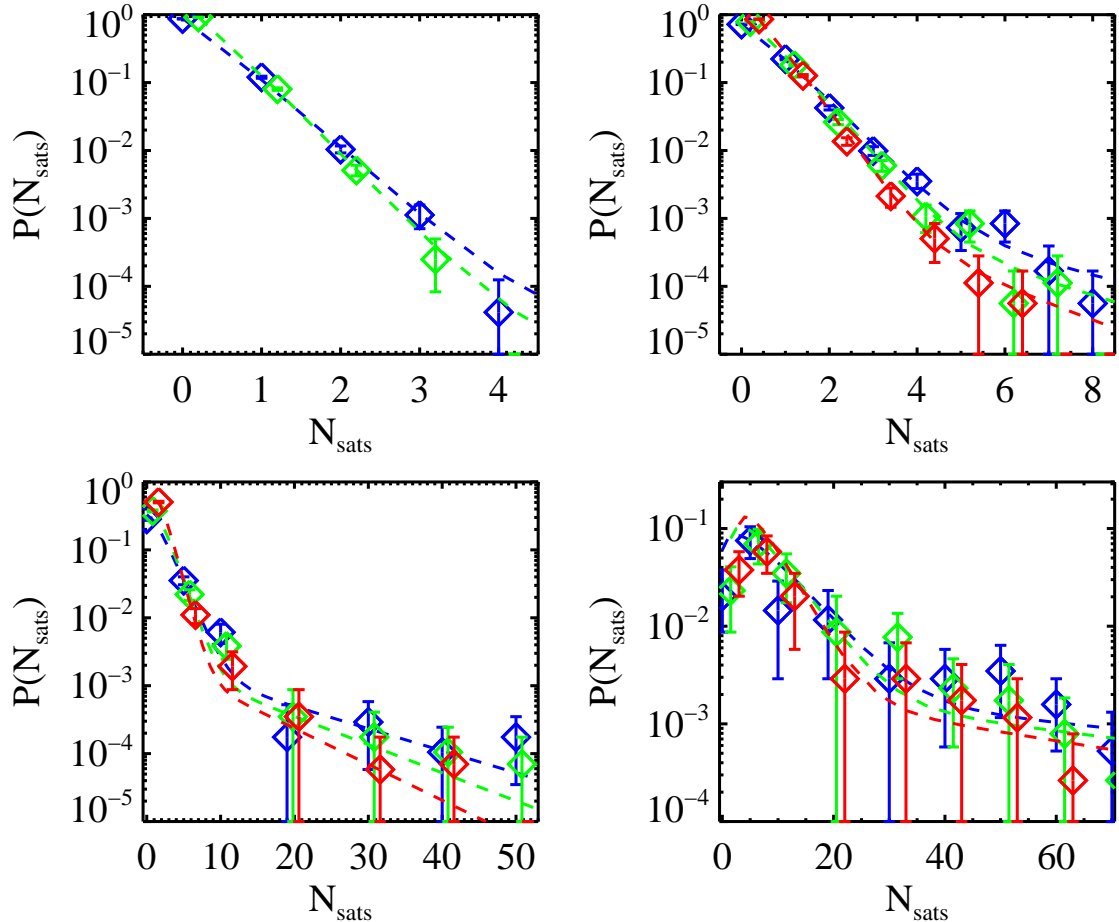} \caption{Probability
    distributions for galaxies to host \NS\ satellites (diamonds).
    The four panels represent different selections based on host
    magnitudes $M_{r,host} = -19.5, -20.5, -21.5, -22.5$ for the top
    left through lower right.  The blue, green, and red points
    represent various limiting satellite magnitudes, $M_{r,lim} =
    -16.5, -17.5, -18.5$.  The dashed lines represent our model with
    fit parameters from Eq. \ref{eq:full_fit} and Table \ref{table:parameters}.}
\label{fig:fit_verification}
\end{figure*}

Guided by these observations, and noting that $\alpha_2$ and $T$ don't
have strong trends with either host magnitude or limiting luminosity
ratio, we propose a 7-parameter fit to generally describe the full
$P(N|M_{r,host}, M_{r,lim})$.  We use our generalized
negative-binomial plus power law as defined in Equation \ref{eq:fit},
and keep $\alpha_2$, $T$, and $\tau$ constant.
However, we parameterize 
$\langle N \rangle$ as a linear function, the slope of which depends
on $M_{r,host} - M_{r,lim}$ 
and the offset of which depends on host magnitude, $M_{r,host}$.  
Our full fit for Equation \ref{eq:fit} then becomes a function of the
parameters $N_{0.a}, N_{0,b}, N_{1,a}, N_{1,b}, \alpha_{2}, T_0$, and
$\tau_0$ defined by: 
\be
P(N|M_{r,host}, M_{r,lim}) = {\Gamma(N+r) \over \Gamma(r)\Gamma(N+1)}
p^r (1-p)^N + T_{0}e^{-\tau_{0} N} 
\label{eq:full_fit}
\ee
where
\bee
p &=&  {1 \over 1 + (\alpha_2^2 - 1) \langle N \rangle (M_{r,host}, M_{r,lim})} \nonumber \\
 r &=& {1 \over \alpha_2^2 - 1}  
\label{eq:parameters}\\
\langle N \rangle  (M_{r,host}, M_{r,lim})&=& -10^{N_{0,a} +
  N_{0,b}*\log(-M_{r,host}+5\log(h))} - \nonumber \\ 
& & (M_{r,host}-M_{r,lim})10^{N_{1,a} + N_{1,b}*\log(-M_{r,host}+5\log(h))}\nonumber.
\eee
We fit this functional form to the measured \PNS\ distribution for all
host halos down to $M_r = -18.5$ and their satellites down to $M_r =
-16$ in Bolshoi stacked in bins of 0.5 mag.  We present our best
fitting parameters for Equation \ref{eq:parameters} 
in table \ref{table:parameters}.  As a final validation of this
result, Figure \ref{fig:fit_verification} shows the measured $P(N)$
distributions for a range of $M_{r,host}$ and $M_{r,lim}$, along
with the fits from our model in Eq. \ref{eq:full_fit}.  \mike{Note that, because the slope and intercept of $\langle N \rangle(M_{r,host},M_{r,lim})$ are linear functions in log space, the exact value of $\langle N \rangle$ is highly sensitive to the parameters $N_{0,a}, N_{0,b}, N_{1,a}$, and $N_{1,b}$.  Because of this, it is necessary to constrain these parameters to 4 significant figures (which is easily done given the constraints from the Bolshoi simulation).}

While our nominal fit can be improved using a $\tau$ that is a
power-law in $M_{r,host} - M_{r,lim}$ with a normalization that
depends on host mass, the actual fit does not significantly improve
the reduced $\chi^2$, so we elect to hold it constant. 

\begin{table}
\begin{center}
\caption{Parameters specifying $P(N)$ distribution in Eq.~\ref{eq:parameters}}
\begin{tabular}{ c  r }
\hline
parameter & best-fit value\\
\hline
\hline
$N_{0,a}$ & -53.70\\
$N_{0,b}$ & 40.11\\
$N_{1,a}$ & -39.05\\
$N_{1,b}$ & 29.25\\
$\alpha_{2}$ & 1.3 \\
$T_0$ & $1.3 \times 10^{-4}$\\
$\tau_0$ & 0.023\\
\hline
\end{tabular}
\end{center}
\label{table:parameters}
\end{table}

\section{Conclusions and Summary}  
\label{sec:conclusions}

In this work, we have examined the likelihood for simulated MW-like
dark matter halos to host substructures similar to the Magellanic
Clouds.  We have performed this analysis using both dark matter
kinematic information and an abundance-matching technique which
assigns galaxy luminosities to resolved dark matter halos and allows
direct comparison with observations (e.g., L10).  The main result of
this work is that MW-like objects have a 5-11\% chance to host two
subhalos as large or as luminous as the SMC, in basic agreement with
previous simulation results (e.g., BK10), as well as with
observations.  While we have extensively compared our results for the
full probability distribution for a MW-luminous galaxy to host \NS\
satellite galaxies to the observational measurements from L10, finding
good agreement.  This results are also in qualitative agreement with
previous results using smaller samples \citep{Chen06, James10}.

Our results have a number of implications.  First, this is a
validation of the \LCDM paradigm down to the level of
$10^{10}\hinv\msol$ objects.  At this level, there is no indication
that either the CDM paradigm or our standard picture of galaxy
formation, including the relatively tight relationship between galaxy
luminosities and halo masses, breaks down.  In particular, it appears
that the statistics of satellite galaxies at this mass in LCDM are in
very good agreement with observations which sample a wide range of
galaxy environments.

One possible criticism of the model we have used for connecting
galaxies to halos (SHAM) is that it may be so robust as to {\it
  always} reproduce the observed statistics, given its assumption of
the correct global luminosity function.  In order to evaluate this, it
is important to keep track of the exact assumptions of the approach
and how these relate to the particular statistics measured.  First and
foremost, SHAM assumes that $\vmax$ of a halo or subhalo (at accretion
into a larger system) is the {\it only} parameter that sets the
luminosity of the galaxy it hosts.  While this is robust to, e.g., a
temporally-variable star formation efficiency with halo mass, it would
differ from model in which there was environmental dependence to the
galaxy formation recipe in addition to the inherent environmental
dependence of the halo mass function.  For example, if subhalos of a
given mass evolve very differently than halos of the same mass in the
field, our model would not produce good agreement.  Additionally, if
our simulated dark matter halo mass function was significantly off
from reality (due to, for example, an incorrect $\sigma_8$ or
$\Omega_m$), the masses of the MCs as estimated from abundance
matching would not provide a good match to the direct kinematic
measurements, as we find here.  Finally, this is a test of the SHAM
treatment of subhalos in a mass regime lower than has been previously
explored.

We do see some manifestations of assembly bias with regard to the dark
matter subhalo population inside MW-like galaxies: halos in higher
density regions host more massive subhalos than those in lower density
regions.  This effect appears to be most pronounced for relatively
local measurements of density, and suggests that the MW's proximity to
M31 boosts the likelihood for us to see a LMC/SMC pair by about 25\%.
The effect is strongest for densities measured on the scale of 1 Mpc.
We note that such a boost seems to be present if we use either the
total local mass or the total local luminosity as an environmental
measure.  This should make the effect an observable manifestation of
assembly bias (though detecting it will require careful treatment of
correlated structure).  It is interesting that this result is in
qualitative agreement with the work of \cite{Ishiyama08}, who saw a
similar trend with environment but claimed it was most sensitive at
the $\sim 5\hinv\mpc$ scale.  They note that we are in an underdense
region on these scales, and cite this as a possible ingredient for
solving the missing satellite problem.  While our simulations do not
allow us to determine the probabilities of finding objects smaller than the
Magellanic Clouds, if our observed trend were extrapolated down to
lower masses it would result in an {\it opposite} effect, where the
presence of M31 could serve to exacerbate the missing satellite
problem.

Finally, we have generalized our measurements for the \PNS\
distribution over a wide range of host and satellite luminosities.
Our results show a significant deviation from the Poisson expectation
for larger values of \NS.  While these deviations are typically
present only in the low-likelihood tail of the distribution, the large
volume of surveys like SDSS should allow tests of these predictions
with observations.  Deviations from Poisson statistics in the number
of satellites may also have consequences for HOD modeling, which
currently assume a Poisson distribution for the number of galaxies
given the halo mass.

\acknowledgments RHW and MTB were supported by the National Science
Foundation under grant NSF AST-0908883.  RHW, PSB, and BFG received
support from the U.S. Department of Energy under contract number
DE-AC02-76SF00515.  AK and JRP were supported by NASA ATP and NSF AST grants.
The Bolshoi simulation was run on the NASA Advanced Supercomputing
(NAS) Pleiades computer at NASA Ames Research Center.  We thank Louie
Strigari, Lulu Liu, and Anna Nierenberg for useful discussions.

\bibliographystyle{apj}

\end{document}